\documentclass[reprint,amsmath,amssymb,aps,prl]{revtex4-2}

\usepackage[utf8]{inputenc}
\usepackage{graphicx}
\graphicspath{{./figures/}{./}}
\usepackage[dvipsnames,x11names]{xcolor}
\usepackage{hyperref}
\usepackage{physics}
\usepackage[capitalize]{cleveref}

\begin{document}

\title{Spatially patterned phases in a reaction-time-symmetry-broken model of flocking}

\author{Charles R. Packard}
\author{Daniel M. Sussman}\email{daniel.m.sussman@emory.edu}
\affiliation{Department of Physics, Emory University, Atlanta, GA, USA}

\date{\today}

\begin{abstract}
We introduce a Vicsek-like flocking model with a minimal form of time-delayed orientational interactions, in which the delays occur on a time scale that is well-separated from other time scales in the model.
We achieve this by implementing an ``index-ordered'' update rule, mimicking a scenario in which agents have a distribution of times with which they react to information.
This model retains the usual disorder-to-order transition common in flocking models, but we show that it also possesses a second transition, deep in the polar flocking phase, to a state with spatially patterned transverse velocities. 
We characterize this transition and its sensitivity to finite-size effects using the Binder cumulant, and demonstrate --- via direct measurements and by measuring a susceptibility of the phase to particle index permutations --- that the stability of this phase is directly tied to a subtle spatial organization of a slow-relaxing index-order field.
These results highlight the potential for even seemingly insignificant temporal asymmetries to fundamentally alter the collective behavior of active matter. 
\end{abstract}

\maketitle


Understanding and quantifying the spontaneous dynamical motions of large groups of living systems --- one step on the way to a theoretical treatment of ``living matter'' --- is a formidable challenge in non-equilibrium statistical physics \cite{marchetti2013hydrodynamics}.
Within and between organisms there are spatial and temporal processes that often have no clear separation of scales, and truly microscopic treatments are doomed to drown in the enormous complexity of these problems.
Pioneering computational work of Reynolds \cite{reynolds1987flocks} and Vicsek \cite{vicsek1995NovelTypePhase} focused not on these complex details and instead introduced radically simplified computational models of ``bird-oid'' objects: self-propelling agents with polar aligning interactions.

These agent-based models were foundational to our understanding of polar active matter and remain cornerstones of active matter physics.
They continue to inspire a substantial body of research investigating new universality classes \cite{jentsch2024NewUniversalityClass}, measuring critical exponents \cite{mahault2019QuantitativeAssessmentToner}, and understanding the role of noise and model structure on the set of non-equilibrium steady states that Vicsek-like models support \cite{martin2021FluctuationInducedPhaseSeparation,ginelli2010RelevanceMetricFreeInteractions,packard2024BandedPhasesTopological}.
The theoretical treatment of these models by Toner and Tu \cite{toner1995LongRangeOrderTwoDimensional,toner1998FlocksHerdsSchools,toner2024physics} was originally intended to describe flocks of  ``real living organisms... provided that they have the same symmetries and conservation laws that, e.g., Vicsek's algorithm does'' \cite{toner2018WhyWalkingEasier}.
While Toner-Tu-like hydrodynamic descriptions can be derived via explicit Boltzmann-like coarse-graining \cite{bertin2006boltzmann,ihle2011kinetic}, it is often more appealing to write down hydrodynamic equations for a natural set of macroscopic fields on the basis of symmetry (including terms that might be forbidden for systems in thermal equilibrium) \cite{marchetti2013hydrodynamics}.

Despite the successes of this theoretical framework in describing some experimental data \cite{geyer2018sounds}, in the absence of explicit coarse-graining of microscopic details it is not always straightforward to enumerate the relevant symmetries appropriate for collective motion of specific biological systems, or even for a specific model~\cite{maitra2025inconvenienttruthflocks,chepizhko2021revisiting}.
Prominent examples of these subtleties include the role of number conservation \cite{toner2012BirthDeathFlight} and of inertia \cite{cavagna2023NaturalSwarms399} on the hydrodynamics of flocking.
In describing actual flocks of birds or other organisms as extremely simplified collections of point particles, we (implicitly or explicitly) imagine coarse-graining over an enormous number of out-of-equilibrium degrees of freedom.
Here it is even less clear which symmetries we can still assume are preserved, or which are broken but in a way which is irrelevant (perhaps corresponding to a fast-relaxing field) in the hydrodynamic limit.
These challenges have motivated a deeper focus on how broken symmetries at the microscopic scale --- such as non-reciprocal interactions \cite{dadhichi2020nonmutual,chen2017fore,mangeat2024emergent,sinha2024reciprocity,fruchart2021non}, or the shift from metric to topological interaction criteria \cite{ballerini2008interaction,ginelli2010RelevanceMetricFreeInteractions,sussman2021NonmetricInteractionRules,martin2021FluctuationInducedPhaseSeparation,packard2024BandedPhasesTopological} --- can profoundly alter phases of collective motion.

One important difference between the interactions in biological flocks and forces in classical mechanics is that, after coarse-graining over non-equilibrium degrees of freedom, the interactions between degrees of freedom in a flock need not be local in time.
For instance, data from flocks and herds show a distribution of delay times associated with the aligning interactions between members of the collective \cite{nagy2010HierarchicalGroupDynamics,gomez2022intermittent}.
Understanding the role of such time-delayed interactions is complex, and a large body of work has attempted to explore the effects of such interactions within the context of Vicsek-like models \cite{nagy2010HierarchicalGroupDynamics,sun2014TimeDelayCan,erban2016cucker,durve2018ActiveParticleCondensation,choi2019hydrodynamic,chen2020ProbabilisticCausalInference,chen2024PersistentResponsiveCollective,holubec2021FiniteSizeScalingEdge,wang2023spontaneous,geiss2022SignalPropagationLinear,pakpour2024DelayinducedPhaseTransitions}.
A major challenge is that models with generic time-delayed interactions can involve arbitrarily complicated memory kernels; in some cases, one even needs to consider how delays themselves vary over time \cite{giuggioli2015delayed,zhang2021stability}.

\begin{figure}[htb]
\centering
\includegraphics[width=0.95\linewidth]{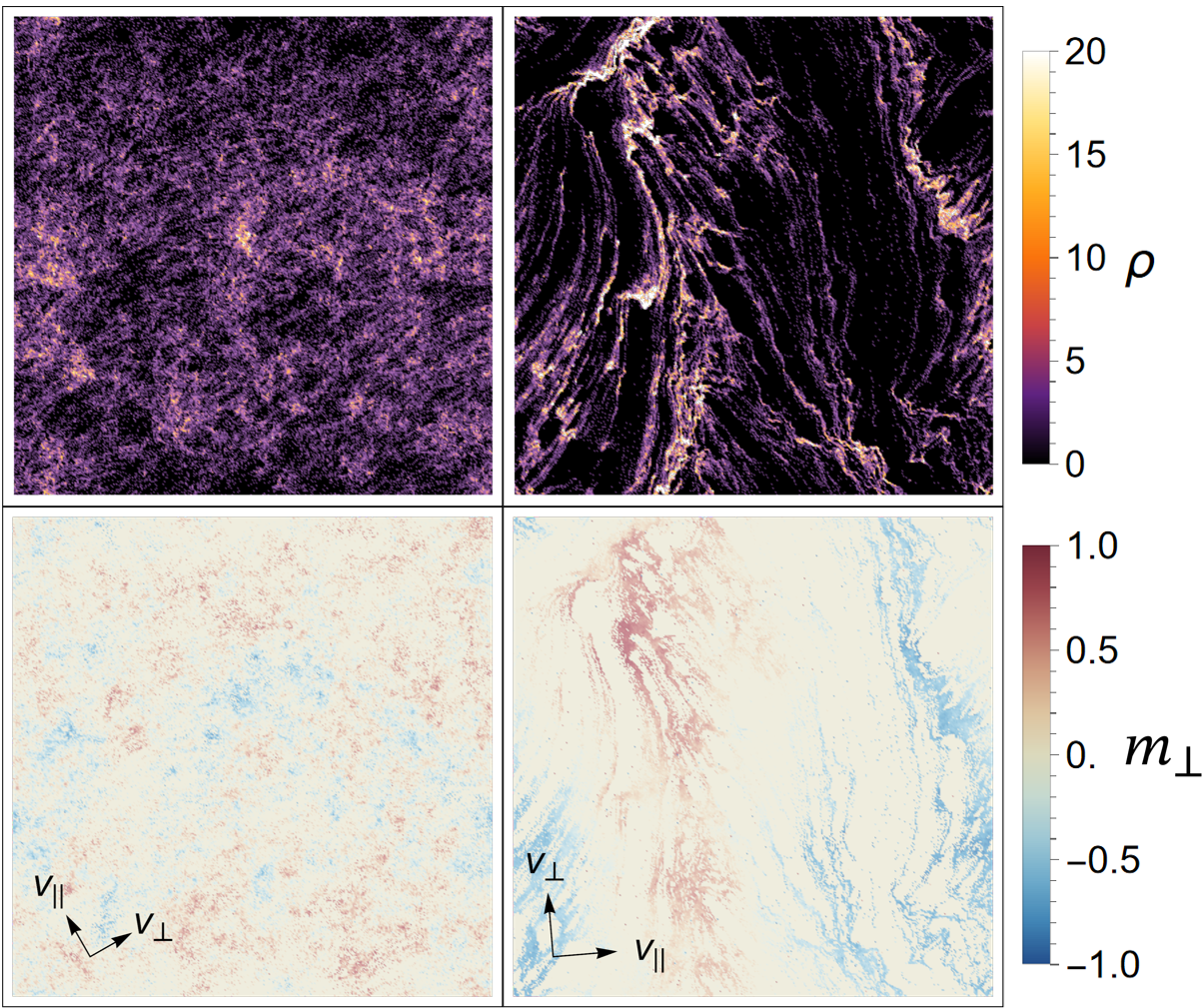}
\caption{
    \textbf{Snapshots of polar and spatially patterned flocking phases.}
    Density (top row) and transverse velocity (bottom row) fields corresponding to snapshots of simulations of $N=2^{17}$ particles at noise strength $\eta = 0.09$ (left column) and $\eta=0.03$ (right column). The fields were computed by averaging over spatial regions of area $\approx 1$ in a simulation domain of side length $L=\sqrt{N}$, and arrows indicate the coordinate frame corresponding to the mean flocking direction ($\textsf{v}_{||}$) in the snapshot.
    }
\label{fig:snapshots}
\end{figure}

This complexity makes it challenging to isolate the fundamental effects of time-delayed interactions; should a small breaking of ``reaction-time'' symmetry in a flock correspond to new hydrodynamic terms?
Should we expect those new terms to give rise to new collective dynamics
in general? 
In this manuscript, we address such questions by introducing a minimal model of Vicsek-like flocking dynamics in which there is a clear separation of time scales in how agents translate and turn on the one hand and process orientational information from their neighbors on the other. 
We demonstrate that by doing so --- breaking the ``reaction-time'' symmetry of the agents \emph{within a numerical timestep} --- we find an additional flocking phase at low noise strengths. 
This flocking phase is characterized by a spatially patterned structure (illustrated in \cref{fig:snapshots}) in which two bands propagate together in the global flocking direction but with opposite transverse velocities.
This highlights the importance of considering even small temporal asymmetries in active matter systems.


We study the following Vicsek-like model of $N$ interacting particles in a 2D square box of linear size $L$ with periodic boundary conditions, written in terms of its computational (discrete time) dynamics.
These particles self-propel at a characteristic speed $v_0$ along their director, which is characterized by an angle in the plane, $\hat{n}(\theta)$:
\begin{equation}
    \mathbf{r}_i(t+\Delta t) = \mathbf{r}_i(t)  + v_0 \Delta t \hat{n}(\theta_i(t)).
\end{equation}
There are numerous variations of the Vicsek model that specify different polar aligning interactions between particles, and we begin by describing an Active-Brownian-Particle-like model for the orientational dynamics:
\begin{equation}
    \theta_i(t+\Delta t) = \Delta t \,\tau_i + \eta\sqrt{\Delta t}\zeta_i(t),
\end{equation}
where $\zeta_i(t)$ are uncorrelated random variables uniformly distributed in $(-\pi,\pi)$, and $\eta$ controls the strength of the fluctuations in $\theta$.
To remove the effects of non-reciprocal ``neighbor-number''-mediated interactions in the original Vicsek model \cite{chepizhko2021revisiting,packard2022non,sinha2024reciprocity}, we first consider particles whose torques are derived from a XY-Hamiltonian: $\tau_i = -\nabla_{\theta_i} H$ for 
\begin{equation}\label{eq:rtsH}
H[\{\theta_{i}(t)\}]=-\alpha\sum_{{j}\in\mathcal{N}_i(t)}\cos\left[\theta_{i}(t)-\theta_{j}(t)\right].
\end{equation}
The parameter $\alpha$ characterizes how fast particles would align in the absence of noise, and $\mathcal{N}_i(t)$ is the set of neighbors of particle $i$ at time $t$.
This set is chosen to be the instantaneous set of Voronoi neighbors of particle $i$, and in order to explore sufficiently large systems in a computationally efficient manner we adapt GPU-accelerated code from the \emph{cellGPU} package \cite{sussman2017CellGPUMassivelyParallel}.
Vicsek-like models are well-known to be sensitive to finite-size effects, and we use this ``topological'' version of the model as we find patterned phases in it at somewhat smaller (and hence, more easily studied) system sizes.
In the Supplemental Material \cite{SupMat} we present evidence that a metric version of the this model displays similar phenomenology.

\Cref{eq:rtsH} reflects a situation in which particles experience equal-time alignment interactions --- the ``reaction-time-symmetric'' Hamiltonian corresponds to torques at time $t$ based on the state of particles at time $t$. 
We wish to break the symmetry of these aligning interactions in a minimal way.
Inspired by the observed distribution of reaction times at the scale of organisms, but not wanting to model the complexity of such full distributions of time-delays in a large flock, we imagine an \emph{index-ordered} set of orientational updates with all asymmetry occurring within a timestep.
Concretely, we consider particle $i$ perceiving ``time-local'' information from neighbors of lower index and time-delayed information from neighbors of higher index:
\begin{equation}\label{eq:rtsbH}
    \tau_i = \alpha\sum_{{j}\in\mathcal{N}_i(t)}
\left\{
   \begin{matrix}
       \sin\left[\theta_{j}(t+\Delta t)-\theta_{i}(t)\right] &  \textrm{if } i > j \\
       \sin\left[\theta_{j}(t)-\theta_{i}(t)\right] & \textrm{if } i < j 
   \end{matrix}
\right..
\end{equation}
One can interpret this as a scenario in which low-index particles experience relatively more time-delays than high-index particles, albeit with the time delays evenly distributed within a single computational timestep.
This reflects time-delays with a clear separation between the time-scale of the delay interaction and the time scale over which particles orient and translate.
Unless otherwise specified, we work with simulations parameterized by $v_0=0.5$, $\alpha=4$, at fixed number density $\rho = N/L^2=1.0$, and with a time-step size of $\Delta t=1$.


We perform simulations of this ``index-ordered time-delay'' model across a range of noise strengths and system sizes. 
We confirmed that this model recapitulates the usual transition between a disordered active gas for $\eta \gtrsim 0.4$ and a polar flocking state below that. 
As we continue to decrease the noise strength, though, we unexpectedly see a transition to the unusual flocking state shown in \cref{fig:snapshots}.
In this state a global flocking direction (denoted by $\textsf{v}_{||}$) remains, but the system spatially organizes into two bands of high density; these bands have large (and opposite) mean values of particle velocity transverse to the global flocking direction ($\textsf{v}_\perp$).

\begin{figure}[b!]
\centering
\includegraphics[width=0.9\linewidth]{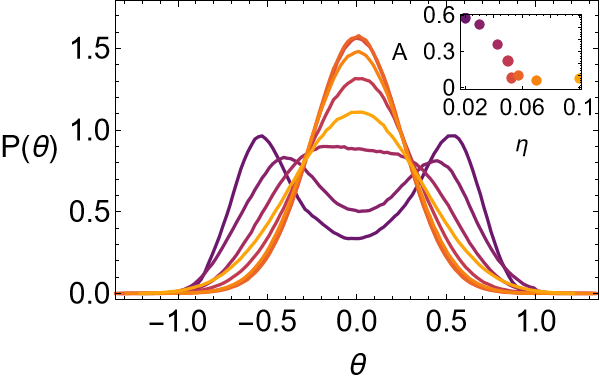}
\includegraphics[width=0.9\linewidth]{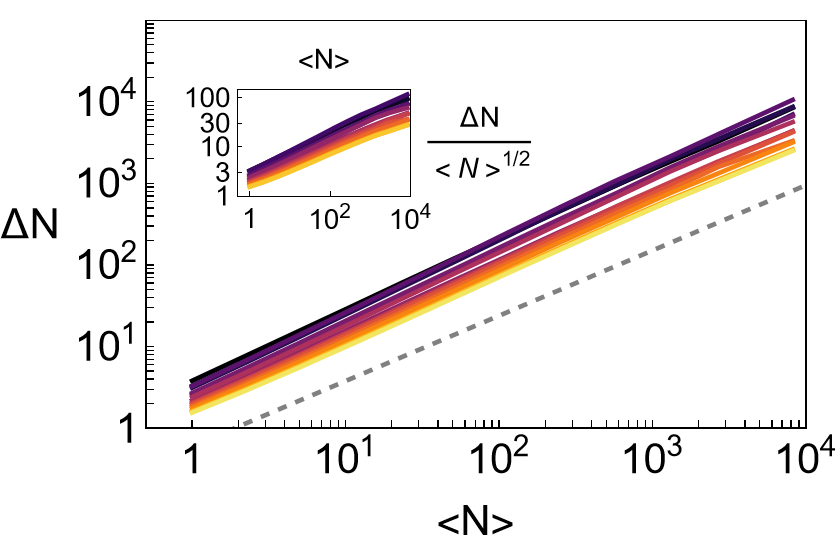}
\caption{
    \textbf{Orientation and density statistics at low noise.}
    Statistics of simulations of $N=2^{17}$ particles, averaged over 40 snapshots each separated by $t=1000$, for (dark to light) $\eta = 0.02, 0.03, 0.0425, 0.05, 0.0525, 0.0575, 0.07, 0.1$.
    (Top) The probability density of angles measured relative to $\hat{v}_{||}$. The inset shows the amplitude of the best-fit sinusoidal profile of $v_\perp$ averaged in bins transverse to the flocking direction as a function of noise strength.
    (Bottom) Number fluctuations $\Delta N = \sqrt{(N-\langle N\rangle)^2}$ vs $\langle N \rangle$; the dashed gray line is a guide to the eye of slope $0.8$.
    The inset shows $\Delta N / \sqrt{\langle N\rangle}$, clearly distinguishing these results from Gaussian number fluctuations.
    }
\label{fig:phaseTransitionSignature}
\end{figure}

To understand the properties of this unusual state --- a banding regime \emph{not} between the disordered and ordered state as in the usual Vicsek model \cite{ginelli2016physics}, but deep within the polar flocking state --- we begin by quantifying the orientational statistics of particles near the transition and the structure of number fluctuations.
The top panel of \cref{fig:phaseTransitionSignature} shows the distribution of particle directions measured relative to the mean flocking direction, $\theta_i = \arcsin\left(\hat{n}_i\cdot \hat{v}_\perp\right)$.
Most notably, as the noise is reduced a clear change is observed from Gaussian distributed relative angles to a bimodal distribution, which is accompanied by a structural transition into an inhomogeneous ``banded phase.'' This is seen in the inset, in which the mean transverse velocity is averaged in bins transverse to the flocking direction, and $A$ is defined as the amplitude of the best-fit sinusoidal profile.
The bottom panel of  \cref{fig:phaseTransitionSignature} reveals that in the spatially patterned phase giant number fluctuations --- another hallmark of the polar flocking state and a clear departure from the expectations in equilibrium systems --- persist, with an exponent consistent with the expectation for the Vicsek universality class \cite{chate2008collective}.

\begin{figure}[t!]
\centering
\includegraphics[width=0.9\linewidth]{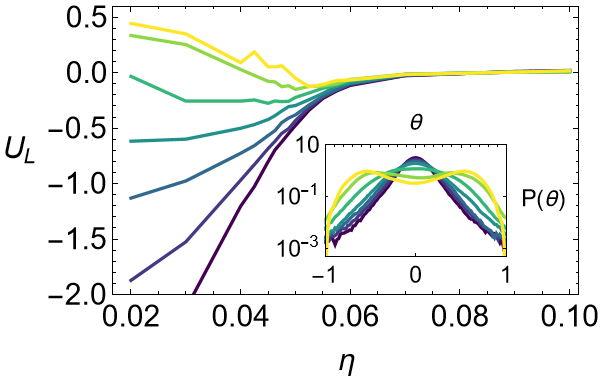}
\caption{
    \textbf{Order parameters for the spatially patterned flocking phase.}
    The main plot shows the Binder cumulant of angles relative to the global flocking direction as a function of noise. System sizes (dark to light) correspond to $N=2^{11} - 2^{17}$ in factors of two. 
    The inset shows the probability density of angles measured relative to $\hat{v}_{||}$ in a log-log representation for the same range of system sizes and for $\eta =0.02$.
    }
\label{fig:finiteSizeScaling}
\end{figure}

Given the bimodal nature of the distributions of angles relative to the global flocking direction, a natural order parameter characterizing the transition is the Binder cumulant, $U_L = 1-\frac{\langle\theta^4\rangle}{3\langle \theta^2\rangle^2}$, which vanishes for Gaussian distributed relative angles.
As shown in \cref{fig:finiteSizeScaling}, we see the breakdown of Gaussian statistics at small $\eta$ for all system sizes $L=\sqrt{N/\rho}$. For small system sizes this manifests itself as a long-tailed distribution and a negative Binder cumulant, and for sufficiently large systems sizes ($N \gtrsim 2^{16}$) the Binder cumulant is positive: clearly bimodal distributions emerge, corresponding to the counter-propagating bands shown in \cref{fig:snapshots}.

Since these spatially patterned phases have not been seen in the standard Vicsek model or its many other variations, it is natural to conclude that they are a direct result of the index-ordered time delays imposed within each timestep.
One possible theoretical treatment would be to posit a coupling term between the usual Toner-Tu equations and the evolution of an additional hydrodynamic field corresponding to the particle index, or perhaps corresponding to the typical relative difference of indices (since relative index ordering is what enters in the microscopic dynamics).
Unfortunately, we find that such a relative index field is prohibitively difficult to measure at levels beyond the noise --- in the Voronoi construction of neighbor lists each particle has on average only six neighbors, and thus there are large discrete jumps in any order parameter that might correspond to gradients in the flocking direction of positive/negative index differences. 

\begin{figure}
\centering
\includegraphics[width=0.9\linewidth]{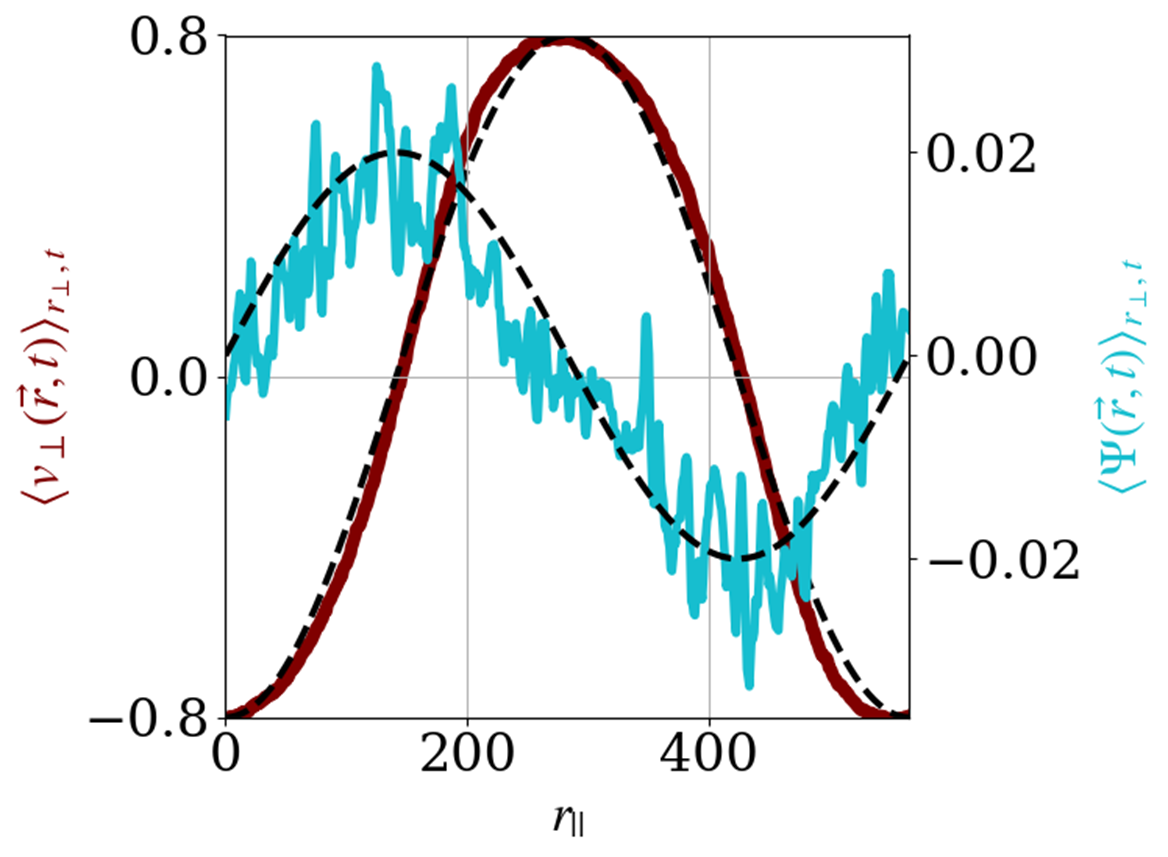}
\caption{
    \textbf{Longitudinal profiles of transverse velocity and torques attributable to time delays.}
    Both the transverse velocity (dark red) and non-reciprocal $\Psi$ (light blue) longitudinal profiles have approximately sinusoidal functional forms, with dashed lines indicating best-fits. Data is presented for simulations with $N=80000$ particles at $\eta=0.01$, averaged over $400$ snapshots each separated by $t = 500$.
    }
\label{fig:forceVelocityProfiles}
\end{figure}

In the Supplemental Material \cite{SupMat} we show that these patterned phases are unstable to instantaneously permuting the indices of the particles in our simulations (which corresponds to measuring an ``index-susceptibility''-like quantity). A more direct demonstration that the pattern phases directly stem from these time-delayed interactions is in \cref{fig:forceVelocityProfiles}. 
There we define $r_{||}$ and $r_\bot$ as the directions parallel and transverse to the flocking direction, and measure the longitudinal profile of fields $f$ averaged along the transverse direction, $\langle f(\vec{r},t)\rangle_{r_\bot}$.
By computing these longitudinal profiles at many time points and averaging the result across snapshots (registering each snapshot by rotating each into the flocking frame and aligning the location of the peaks of the density bands), we find a nearly sinusoidal profile for the mean transverse velocity (corresponding to measurements of $A$ in the inset of \cref{fig:phaseTransitionSignature}).

We then compute a quantity $\Psi$, the difference between the torques computed as in \cref{eq:rtsbH} and the torques that would have been computed via the time-local Hamiltonian in \cref{eq:rtsH}, for each configuration.
Whereas \cref{eq:rtsH} corresponds to standard interactions, the time asymmetry in \cref{eq:rtsbH} departs from the usual action-reaction symmetry of Newton's laws.
The field $\Psi$ is thus a measure of the average non-reciprocal torques acting on the system due to the time delays in the model.
As shown in \cref{fig:forceVelocityProfiles}, the longitudinal profile of $\Psi$ is well-described by a sinusoidal profile, and which is out of phase with the longitudinal profile of the transverse velocities.
In the absence of this coarse-grained non-reciprocal field, particles at the interface of the two counter-propagating bands would align with one another, thereby causing the system to relax into a homogeneous flocking phase.
When $\Psi$ is non-zero though, the system is continuously driven by torques acting uniformly across the interfaces of the two bands.
This is particularly fascinating because while we do impose a time-ordering on the system, $\Psi$ arises from a spontaneous spatial ordering of gradients in the particle index field.
Our measurement of the $\Psi$ field in \cref{fig:forceVelocityProfiles} reinforces the connection between time-delayed interactions and time-local but non-reciprocal interactions \cite{loos2019fokker,loos2021stochastic}, although we are not able to find an analytical mapping in this vein that corresponds to our precise time-delayed interaction scheme.
This may be a fruitful avenue of further study, as the kinds of anomalous advection terms expected at the hydrodynamic level when Newton's third law is evaded may provide an alternative to trying to formulate a continuum description in terms of couplings to hard-to-measure index fields.

In summary, we have introduced a minimal model in which the reaction-time symmetry of agents in a Vicsek-like model is broken, leading to the emergence of a spatially patterned flocking phase.
The existence of this phase in a regime in which all time delays occur \emph{within} a single computational timestep highlights how subtle the construction of hydrodynamic descriptions can be in out-of-equilibrium systems. 
Although we expected an ``index-asymmetry'' field to be a fast-relaxing mode, we nevertheless (indirectly) find that it maintains structure over long times and leads to spatial patterning of the density and velocity fields. 
We note that infinitesimal time delays in oscillating systems can be interpreted as negative effective damping terms \cite{jenkins2013Selfoscillation}.
This combined with a globally favored flocking direction at low noise may provide a rationalization of the structure of the phase we observe, although there is no clear mechanism by which this would lead to spatial structure in the index field.

In this work we have focused on a version of \cref{eq:rtsbH} in which neighbors are chosen according to a Voronoi construction.
We note that this choice of topological interactions is a convenience --- we observe similar but much more spatially complex patterned phases when we instead use a simpler distance-based criterion for determining neighbors; initial results to this effect are presented in the Supplemental Material \cite{SupMat}, and further work will be reported in a follow-up study.
We presented our construction of \cref{eq:rtsbH} as a minimal model of time delays reflecting (a) the fact that in organisms reaction times are continuously distributed and (b) a separation of time scales between the reaction time of an organism and other time scales of that organism's motion.
This may alternately be viewed as a model with $N$ different species (each possessing an orientational time-delay offset from another species by at least $\Delta t/N$).
In light of this, it would be interesting to study other models of time-delays within a similar class. 
This could include models in which the time-delays have a non-uniform distribution, or models with only a small number of different species --- e.g., one in which one species experiences a $\Delta t$ time delay and the other does not.

\onecolumngrid
\section*{Supplemental material}
\twocolumngrid

In the first section of this Supplemental Material we present the results of measuring a susceptibility-like quantity in the reaction-time-symmetry-broken model presented in the main text. In the second section we present evidence that a metric rather than topological version of the same model has a similar, if more complex, transition to a spatially patterned phase. 

\section{Sensitivity to index permutations}\label{sec:susceptibility}

\begin{figure}[htb]
\centering
\includegraphics[width=0.5\linewidth]{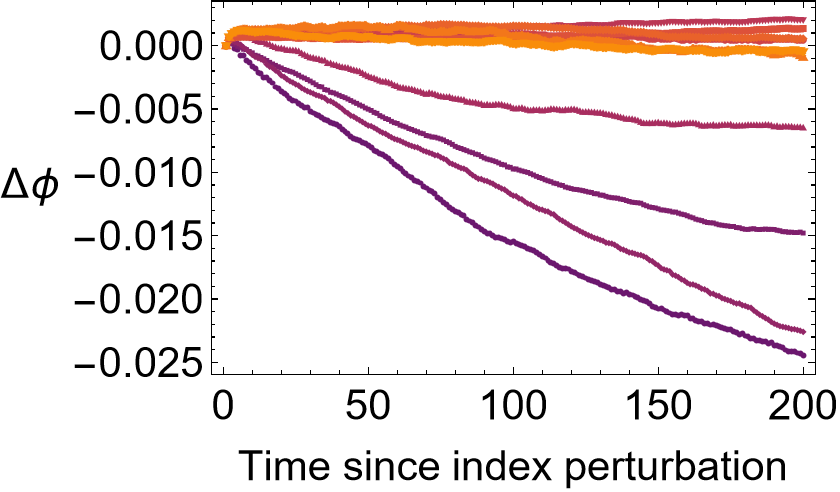}
\caption{
    \textbf{Susceptibility to index permutations}
    We measure the average change in the $\phi$ relative to the original simulation for an ensemble of 50 isoconfigurational copies of the system which each receive a different random permutation of particle indices.
    Data corresponds to simulations of $N=80000$ particles with (dark to light) $\eta = 0.02, 0.03, 0.04, 0.045, 0.0475, 0.05, 0.055, 0.07, 0.09, 0.1$, averaged over 50-200 ensembles per noise strength.
    }
\label{fig:indexPerturbation}
\end{figure}

We have argued that even though the appropriate gradients of a hydrodynamic  ``index field'' is extremely  difficult to measure, the spatially patterned phases observed are the direct result of the index-time-ordering of how torques are computed. As an alternate demonstration of this, we measure an ``index-susceptibility''-like quantity.
We first allow simulations at different noise strengths to reach a steady state, and then apply a random permutation to the particle indices.
That is: we generate a random permutation of the indices from 1 to $N$, and shuffle the order of all particle-level data structures (positions, orientations, etc.) according to this permutation. We then continue to simulate the system with these new particle indices.
We repeat this operation over an ensemble of 50 isoconfigurational copies of the system (i.e., configurations that have the same positions and velocities, each of which are subject to a different random permutation of the indices), and compare with the dynamics of these copies with a simulation in which the identity permutation is applied.
In the usual polar flocking phase this permutation should have no effect, as in that phase we expect no spatial organization of the index field.
In the spatially patterned phase, though, any correlations between the gradients of the index ordering and the direction of flocking would be destroyed by this operation.
We would expect, then, that the torques responsible for maintaining the two counter-propagating bands would be disrupted, leading each band to more strongly order with itself (and, thus, transiently decreasing the \emph{global} flocking order parameter as the system transiently orders into two bands each flocking in a different direction).
As shown in \cref{fig:indexPerturbation}, which measures the change in the global flocking order parameter, $\phi = \frac{1}{N}\sum_{i=1}^N \hat{n}_i$, for an index-shuffled simulation relative to a simulation with the same configuration that was not index-shuffled, precisely this behavior is observed.

\section{Simulations with distance-based neighbor lists}\label{sec:metric}

In the main text we considered a version of the flocking model in which $\mathcal{N}_i(t)$, the set of neighbors of particle $i$ at time $t$, was chosen based on an instantaneous Voronoi tessellation of the current configuration of the points. 
Vicsek-like models are notoriously sensitive to finite-size effects, often requiring extremely large simulations to access their limiting behavior \cite{chate2008collective}, and there are some indications that topological models are typically modestly less sensitive to finite-size effects \cite{ginelli2010RelevanceMetricFreeInteractions}.
Nevertheless, one may wonder whether the unusual spatially patterned phase is a consequence of using a topological determination of a particle's set of neighbors, or if these phases occur more generally in models with this kind of time delay.
To address this question, we have simulated a small number of ``metric'' simulations, in which $\mathcal{N}_i(t)$ is instead the set of all particles within a unit distance of the target particle.
Using metric interactions implies the existence of a new microscopic length scale in the system, and to be concrete we set the particle density 
interactions to be $\rho=2.0$ (so that in the disordered phase a particle would interact with $\approx 2 \pi$ neighbors, a value close to the average of six neighbors per particle in the topological simulations).

\begin{figure}[htb]
\centering
\includegraphics[width=0.95\linewidth]{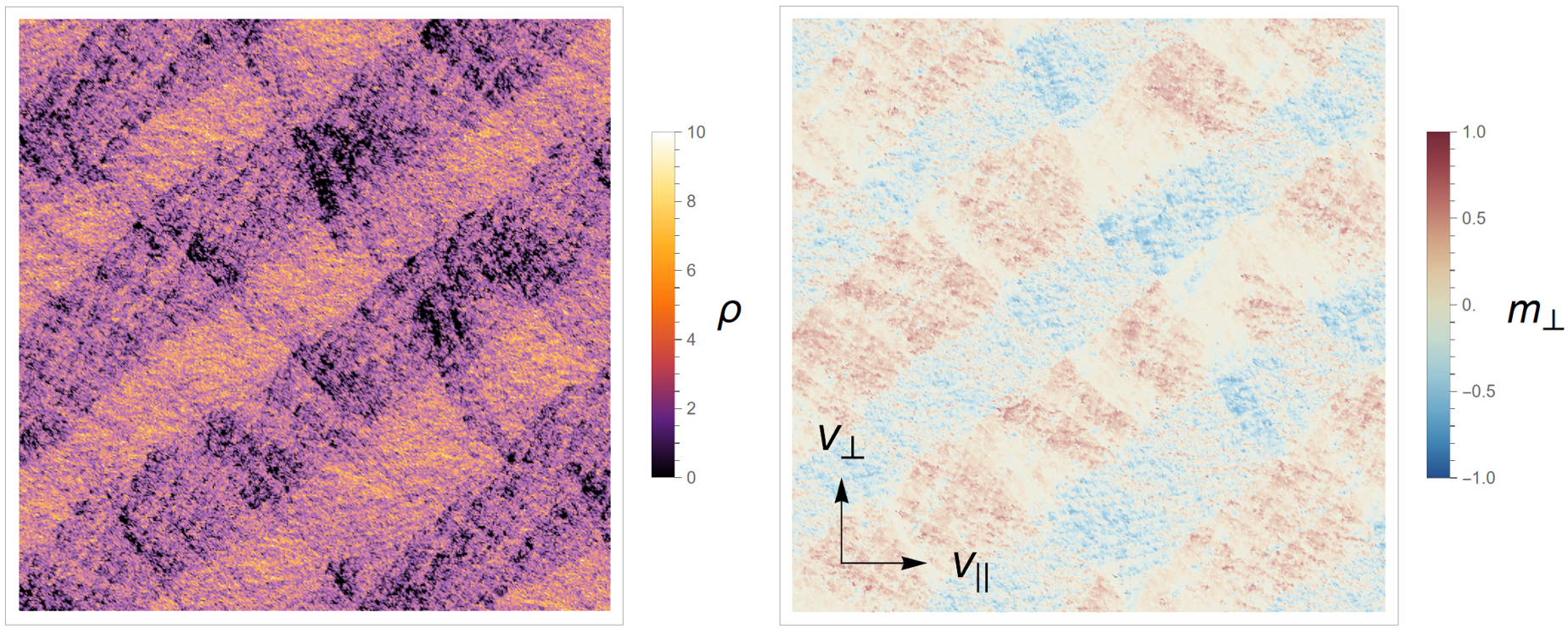}
\caption{
    \textbf{Snapshots from a spatially patterned flocking phase in a metric model.}
    Density (left) and transverse velocity (right) fields corresponding to snapshots of simulations of $N=8\times 10^5$ particles whose set of neighbors corresponds to all particles within a unit distance. The simulations were carried out at density $\rho = 2.0$,  noise strength $\eta = 0.01$, $v_0 = 2.0$, and $\alpha = 0.25$. The fields were computed by partitioning the unit cell into a square grid with cell side length $\approx 1.5$.
    }
\label{fig:metricSnapshots}
\end{figure}

As seen in \cref{fig:metricSnapshots}, at very low noise strength the system transitions from a polar flocking phase to a banding phase, but an unusual one in which the dense \emph{and} dilute regions are each highly ordered, with the dilute regions flocking in a different direction than the dense regions. Observing these unusual metric phases seems to require much larger simulations --- by roughly a full order of magnitude --- than the spatially patterned phases in the Voronoi model reported in the main text. Given the extra computational effort needed, a more thorough investigation of their properties will be reported in a follow-up work.

\begin{acknowledgments}

This material is based upon work supported by the National Science Foundation under Grant No.~DMR-2143815. This research was supported in part by grant NSF PHY-2309135 to the Kavli Institute for Theoretical Physics (KITP). We thank Ilya Nemenman, Randall Kamien, and Helen Ansell for stimulating discussions.
\end{acknowledgments}

\bibliography{bibliographyTimeOrderedVicsek}

\end{document}